\documentclass[letterpaper,12pt]{article}
\usepackage{fullpage}
\usepackage{fancyvrb,fancyhdr}
\usepackage{amsfonts}
\usepackage{array}
\usepackage{caption}
\usepackage{multirow}
\usepackage{amsmath, amssymb, amsthm}
\usepackage{graphicx,longtable, booktabs}
\usepackage[flushleft]{threeparttable}
\usepackage[footnotesize,center]{subfigure}
\usepackage{enumerate}
\usepackage{mathrsfs}
\usepackage{cancel}
\usepackage{soul}
\usepackage{MnSymbol}
\usepackage{algorithm}
\usepackage{algorithmic}
\usepackage{color}

\newfont{\bssten}{cmssbx10}
\newfont{\bssnine}{cmssbx10 scaled 900}
\newfont{\bssdoz}{cmssbx10 scaled 1200}

\newfont{\sserifn}{cmssbx10 at 11pt}
\newfont{\sserifo}{cmssbx10 at 12pt}

\newcommand{\vect}[1]{\mathbf{#1}}

\newcommand{\figref}[1]{Figure~\ref{#1}}
\newcommand{\tableref}[1]{Table~\ref{#1}}
\newcommand{\eqsref}[1]{Eq.~\ref{#1}}

\newcommand{\nst}{N$\sigma$T }

\usepackage[para,multiple]{footmisc}

\begin{document}


\title{Method and Advantages of Genetic Algorithms in Parameterization of Interatomic Potentials: Metal-Oxides}
\author{Jos\'e Solomon $\dagger$ \\ \textit{Mechanical Engineering Department}, Stanford University, CA \\ \ \\  Peter Chung \\ \textit{US Army Research Laboratory}, Aberdeen Proving Ground, MD \\ \ \\ Deepak Srivastava $\ddagger$ \\ \textit{NASA Ames Research Center}, Moffett Field, CA \\ \ \\
 \ \ \  Eric Darve \\ \textit{Mechanical Engineering Department}, Stanford University, CA}
\maketitle

\thanks{$\dagger$ jose.e.solomon@gmail.com} \\
\thanks{$\ddagger$ Initial stages of this work were completed while still a Senior Scientist and Task-Manager at NASA Ames. }

\begin{abstract}
The method and the advantages of an evolutionary computing based approach using a steady state genetic algorithm (GA) for the parameterization of interatomic potentials for metal oxides within the shell model framework are developed and described. We show that the GA based methodology for the parameterization of interatomic force field functions is capable of (a) simultaneous optimization of the multiple phases or properties of a material in a single run, (b) facilitates the incremental re-optimization of the whole system as more data is available for either additional phases or material properties not included in previous runs, and (c) successful global optimization in the presence of multiple local minima in the parameter space. As an example, we apply the method towards simultaneous optimization of four distinct crystalline phases of Barium Titanate (BaTiO$_3$ or BTO) using an ab initio density functional theory (DFT) based reference dataset. We find that the optimized force field function is capable of the prediction of the two phases not used in the optimization procedure, and that many derived physical properties such as the equilibrium lattice constants, unit cell volume, elastic properties, coefficient of thermal expansion, and average electronic polarization are in good agreement with the experimental results available from the literature. \\ \ \\
\begin{small}
\end{small}
\end{abstract}


\section{Introduction}

The temperature and pressure dependent atomistic molecular dynamics (MD) simulations of materials systems lie at the core of many advances in the discovery and optimization of new and novel materials in a wide variety of applications areas. The accuracy of the materials properties resulting from MD simulations, however, is always directly correlated to the quality of the interatomic interaction potentials or force field functions underlying the dynamics. The potential function is comprised of a functional form and an accompanying set of numerical coefficients or parameters fitted to optimize the physical properties of the simulated materials. The functional form is generally an ansatz chosen so as to reproduce known symmetries, crystal structure, and basic physical or chemical properties of the material. An initial dataset comprising of  the crystal structures including the lattices constants, bond lengths and angles, and systematic changes in the energies of the system as these values are changed (in other words elastic constants) are obtained either through higher accuracy ab initio quantum mechanical simulations or through experimental measurements. Given such a reference dataset and a chosen ansatz for the functional form, the potential parameters are generally obtained by finding a set that gives the best fit to a defined set of properties or material features \cite{AllenM.P.Tildesley1987}. Optimizing the parameters for a given set of the properties is often a combination of local curve fitting to the selected materials characteristic trends followed by closed-loop MD or static structural simulations in which known macro-scale materials quantities are targeted by iterative manual adjustment of the chosen parameter sets \cite{Shimada2008}.

The above process by definition is sequential in nature and needs to be wholly repeated whenever a reference dataset is augmented with new materials' structural or properties data. As a result, in the earlier days, for well known semiconductors like silicon more than 30 interatomic potentials or force field functions were developed \cite{Clancy1993} but mainly the Stillinger-Weber or Tersoff potentials for silicon or their derivatives have been used extensively over the last two decades \cite{ Stillinger1985, Tersoff1986, Tersoff1988}. Similarly, for reactive hydrocarbons the original Brenner potential \cite{Brenner1990} was developed in early 90s and has been used extensively over the last two decades \cite{Huang2006}, but only minor changes or improvements have been added since then \cite{Srivastava2002}.

The conventional local gradient based curve fitting methods, such as steepest descent and conjugate gradient, used in the traditional optimization loop: (i) are inherently local in nature and strongly dependent on the initial or starting configurations for the fitting, (ii) have a tendency to get trapped in or find mainly the nearest local minima in a many dimensional fitting parameter space, (iii) can not include incremental addition to the reference dataset or the physical properties without sacrificing completely the previous optimization and starting all over again, and (iv) are limited overall by how much data or how many physical properties can be accommodated for a given set of parameters in the optimization procedure. In certain cases when these methods are used, the limitations of a resulting parameterization are a product of the functional form of the potential itself, but many times these limitations result from the sequential nature and local dependency of the  optimization loop used by local gradient based techniques. For example, in many cases, the commonly used functional forms have been reported with one set of parameterization for the bulk properties and another set of the parameterization for the cluster or surface properties of the same material.

Concurrently, in recent years there has also been a focus on the generation, sharing and availability of more and more materials structure and properties data through focused programs such as Materials Genome Initiative (MGI) \cite{MGI} and broad availability of high-performance computing resources to the materials academic and industrial community. Such data includes not only the quantum mechanical ab initio density functional theory based simulations of wide variety of materials structures and properties but also the synthesis and atomic scale characterization of the structure and properties of the same materials in experiments. It is natural, therefore, that alternate optimization or fitting procedures be investigated for not only the direct search and discovery of new materials and properties from these large databases, but also for the development of the down-the-stream simulation techniques such as MD or MC methods which are explicitly geared towards exploiting the large amount of available data in the literature and online databases.

In this work, we investigate an evolutionary computing (EC) based Genetic Algorithm (GA) method for the optimization of interatomic potentials for molecular dynamics of materials systems. The optimization occurs for a cost function that measures the fitness of the force field parameter set. The optimization or fitness within the GA context is a measure of the agreement between the target values of a reference database, irrespective of the number of structures and properties in the database, and the values produced from a given initial or fitted parameter set. Roughly speaking, the challenges faced by optimization approaches is that, for even a limited range of parameter variations, the fitness function may live on a very rugged landscape often with many closely-spaced local minima.

The  general  GA method  is derived  from biological genetic theory  and  Darwinian evolutionary  principles \cite{Back1996, Eiben2007, Wang2001}.  In the  GA method,  a  randomly  generated  set of parameters, which  constitute the so-called population  pool,  are  adjusted  through  recombination and mutation operators \cite{Haupt2004} to  improve  the  parameter sets  according  to  a metric,  also known as the  fitness. A reference dataset of the atomic configurations and their energies obtained from quantum simulations are stored and the GA subsequently refers to the reference dataset continuously during the fitting procedure.  A randomly generated population of parameter sets is initially created, and the GA then begins the stochastic process of recombining and mutating the population  pool to iteratively generate  better and better evolved parameter sets over subsequent generations which are in agreement with the entire reference dataset.

GA-based evolutionary techniques have been used for a wide range of applications, including the optimization of gas transmission lines in petroleum piping \cite{Goldberg1987}, satellite scaffolding design \cite{Eiben2007}, pattern recognition and image analysis \cite{Bhandarkar1999}, microcircuit design \cite{Liserre2004}, and drug design \cite{Douguet2000} for pharmaceutical applications. Within the context of the development and optimization of force field functions, Wang and Kollman \cite{Wang2001} were one of the first to use a GA method to create a parameterization for an interatomic potential consisting of bonded and non-bonded force field components for non-reactive bio-molecular systems. The implementation presented here is derived directly from Globus et al \cite{Globus1985, Algorithm}, where GA was employed to reproduce the Stillinger-Weber potential parameters for silicon \cite{Stillinger1985} as an initial test case.

In this work we develop and extend the method to the optimization of parameters for interatomic potentials for metal-oxides, which are known to exist in multiple phases at different temperatures. As an example, we apply the GA method to the parameterization of Barium Titanate (BaTiO$_3$ or BTO), for which a reference dataset is first created using a DFT method, and then we explicitly show that it is possible to (i) optimize the multiple phases of BTO in a simultaneous single run, (ii) include incrementally more and more data in the reference dataset during the optimization loop, and (iii) include both the near-equilibrium and far-from-equilibrium configurations in the optimization. All four known phases of BTO, rhombohedral, orthorhombic, tetragonal and cubic, are simultaneously fitted within a single GA parameterization run, and are found to be in good agreement with the entire DFT reference dataset, as well as with the available experimental measurements of the basic mechanical and thermal properties of BTO reported in the literature.

In Section 2, we describe the details of the GA method as used for the parameterization of interatomic force field functions with an example application for the optimization of the shell model \cite{Dick1958} potential for BTO described in Section 3. In Section 4, the main results of the GA fitting to multiple phases of BTO in a single simultaneous run are reported, and the validation of the evolved parameterization in comparison with experimentally reported values of the basic structural, mechanical, and thermal properties of BTO are described. Finally in Section 5, we summarize the main advantages and suitability of the GA method in fitting parameters of force field functions as larger and larger reference datasets for different materials are developed and made available online. 

\section{The Genetic Algorithm (GA) Method for Parameterization of Force Field Functions}
\label{GA_section}

\begin{figure}[htbp]
\begin{center}
\includegraphics[scale=0.75]{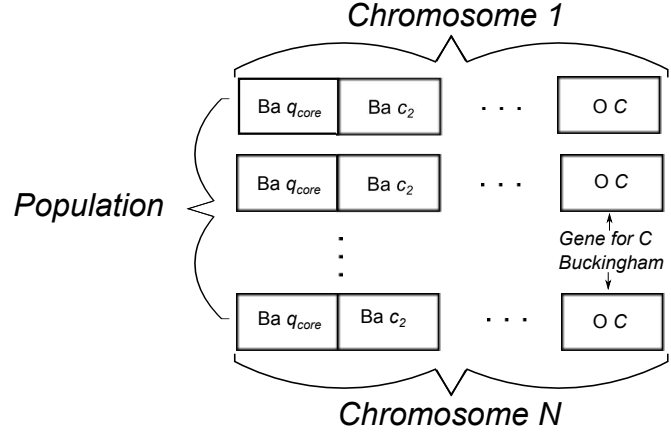}
\caption{Gene, chromosome, and population representations in the GA}
\label{GA_concepts}
\end{center}
\end{figure}

The description of the technique is derived loosely from biological genetic concepts in which it has been established that chromosomes and their respective sequence of genes form the foundation on which heredity leads to environmental adaption through the genotype-to-phenotype cycle. In the GA model a set of genes forms a chromosome and a set of chromosomes forms a population. The concepts of genes, chromosomes and population are illustrated in \figref{GA_concepts}. Each gene corresponds to one parameter in the force field function.  There are as many genes in a chromosome as there are needed to form a complete set of force field parameters.  Thus, a single parameter set corresponds to one complete chromosome but a population may have an arbitrary number of chromosomes.  The technique starts by randomly creating a population of a selected size by generating each chromosome with randomly generated genes. 

The algorithm randomly generates each gene between predefined regime limits specific to that gene. The gene is generated by either sampling from a purely uniform distribution, or from a distribution which is biased toward either end of the gene's parameter limits using a logarithmic distribution \cite{Solomon2012}. A random coin flip decides from which distribution the gene will be selected. Describing a distinct chromosome, we have
\begin{equation}
 \mu_i \in \kappa, \, \mu_i \equiv \{ \eta_1,\eta_2,...,\eta_n \}
\label{chrom}
\end{equation}
where $\kappa$ is the total population, $\eta$ are the distinct genes within the chromosome, $n$ is the total number of genes, and $i$ is the index of the chromosome in the total population pool. The gene generation operator is defined as
\begin{equation}
\hat G(\eta_j,p)  = 
\begin{cases}
\; \eta_j = \text{RandomLogarithmic}(\eta_{min},\eta_{max}) 
& \quad \text{if $p \, \le \, \frac{1}{2}$}
\\ 
\; \eta_j = \text{RandomUniform}(\eta_{min}, \eta_{max})
& \quad \text{otherwise}
\end{cases}
\label{G_op}
\end{equation}
where $p$ is a uniform random number in $[0,1]$ generated at each gene initiation.

RandomLogarithmic($\eta_{min}$,$\eta_{max}$) is the probability function defined by the density
\begin{equation}
P(\eta) = \frac{\frac{1}{X_\text{min or max}}}{\ln (\eta_{max}) - \ln (\eta_{min})} \label{r_log}
\end{equation}
with $X_\text{min or max}$
\begin{equation}
X_\text{min or max} \, = 
\begin{cases}
\; \eta_{max} - \eta & \text{or \ \ \ \ \ \ (a random choice)} \\
\; \eta - \eta_{min}
\end{cases}
\label{X} 
\end{equation}
indicating that the gene is randomly biased toward the maximum or minimum of the gene's parameter limit.

The GA generates a predefined number of population members and this population size is maintained throughout the evolutionary cycle. Thus this variant of evolutionary strategy is characterized as steady state \cite{Globus1985}. To move the population pool towards increasing fitness, a tournament selection operator is used to randomly select two parents from the population pool, 
\begin{equation}
\hat T(\kappa) \rightarrow \{\mu_{\text{mom}},\mu_{\text{dad}} \}
\label{selection}
\end{equation}

Based on this operator, a \textit{child} or offspring is created from its two parents via the recombination operator. In the current implementation, a random selection is made between the parents' gene values, such that
\begin{equation}
 \mu_{\text{child}} \, = \, \hat R(\mu_{\text{mom}},\mu_{\text{dad}}) \label{mu_child}
\end{equation}
where gene $j$ of $\mu_{\text{child}}$ is generated using
\begin{equation}
	\text{RandomUniform}((\eta_{\text{mom}})_j, (\eta_{\text{dad}})_j) \qquad \forall \;  j \in \{1,\dots,n \} \label{recombine}
\end{equation}
(the interval $[(\eta_{\text{mom}})_j, (\eta_{\text{dad}})_j)]$ is used, assuming $(\eta_{\text{mom}})_j \le (\eta_{\text{dad}})_j$) and thus a novel gene set results.

Subsequent to the recombination operator, this new offspring chromosome can be submitted to a mutation operator, which alters the value of a distinct gene based on a predefined mutation frequency $p_m$. For each gene in the chromosome, the mutation operator is
\begin{equation}
\hat M(\eta_i,p,p_m)_{child}  = 
\begin{cases}
\; \eta_j = \text{RandomNormal}(\eta_\text{MF},\sigma) & \text{if $p \, \le \, p_m$} \\  
\; \text{Do not alter} & \text{otherwise}	
\end{cases}
\label{mutator}
\end{equation}
where RandomNormal is a random normal variable with mean $\eta_\text{MF}$ and standard deviation $\sigma$. The value of $\eta_{\text{MF}}$ is taken from the most fit of either of the two parents.

The next step in the cycle is to use an anti-tournament selection operator, in which the least fit of two parents is compared to the offspring. If the offspring is of greater fitness than the weaker parent, this parent is removed from the general population and replaced with the child. In this Darwinian-derived selection process, the population pool is continuously driven toward a more fit genotype. Once a child making cycle is completed, the process is initiated once again and continues until a certain predefined number of generations is produced.  An outline of the method as a whole is given in Algorithm~\ref{GA_algorithm}.

\begin{algorithm}[htbp]
\begin{center}
\caption{Steady state with tournament/anti-tournament selection \label{GA_algorithm}}
\begin{algorithmic}
\COMMENT{Initiate the population}
\FOR{$i = 1:N$}
\STATE \COMMENT{Loop over the genes in a chromosomes}
\FOR{$j =  1:n$} 
\STATE \COMMENT{Random generation of $p$}
\IF{$p \le \frac{1}{2}$}
\STATE $\eta_j = \text{RandomLogarithmic}(\eta_{min},\eta_{max})  $
\ELSE 
	\STATE $\eta_j = \text{RandomUniform}(\eta_{min}, \eta_{max})$
\ENDIF
\STATE Determine resulting chromosome's fitness.
\ENDFOR
\ENDFOR
\STATE \COMMENT{Finished creation of initial population pool $\kappa$}
\STATE \COMMENT{Begin recombination/mutation operators for children}
\WHILE{Generation $\le$ Max Generations}
\FOR{$i = 1:N_{children}$}
	\STATE \COMMENT{Use tournament to select parents from the population pool}
	\STATE $\hat T(\kappa) \rightarrow \{\mu_{\alpha},\mu_{\beta} \}$
	\FOR{$j =  1:n$} 
		\STATE \COMMENT{Perform a recombination operation to produce child genes}
		\STATE $\mu_{child} \, = \, \hat R(\mu_{\text{mom}},\mu_{\text{dad}})$
		\IF{$p \, \le \, p_m $}
			\STATE \COMMENT{Perform a mutation operation on the child's gene if under a predefined mutation frequency}
			\STATE $\hat M(\eta_j,p,p_m)_{child} \, = \, \text{RandomNormal}(\eta_\text{MF},\sigma) $
		\ENDIF
	\ENDFOR
	\STATE Determine resulting child chromosome's fitness.
	\STATE Perform an anti-tournament operator to determine the least fit member of the three chromosomes and update the population pool.
\ENDFOR
\STATE Determine statistics of resulting population pool.
\ENDWHILE
\end{algorithmic}
\begin{algorithmic}
\textbf{-----------------------} \\
\textbf{Legend:}\\
\setlength{\tabcolsep}{2pt}
\begin{tabular}{lll}
$\eta$ &:& gene/parameter value \\
$\mu$  &:& chromosome/parameter set \\ 
$\kappa$ &:& population pool/total set of parameterizations
\end{tabular}
\end{algorithmic}
\end{center}
\end{algorithm}

There are five critical evolutionary parameters that play a prominent role in the evolutionary process which are the size of the population, the number of children-per-generation to be produced, the mutation frequency, the standard deviation of the Gaussian distribution for the mutation operations, and the number of independent GA evolutions that are run based on the exact same parameters. Given the stochastic nature of the GA process and the high-dimensional solution space of the problem, a series of initial trial-and-error evolutions were conducted that provided a range of combinations of these parameters which proved to be the most productive. The results presented in this work are based on the evolutionary parameters listed in \tableref{GA_params}

\begin{table}[htbp]
\begin{center}
\caption{Parameters of the Evolutionary Cycle}
\begin{tabular}{|l| c |}
 \hline
Evolutionary Parameter & Value  \\ \hline
Population Size & 400 \\ 
Child-per-Generation & 160 \\ 
Mutation Frequency & 75$\%$ \\
STD of Mutator & 2 \\
Number of Concurrent Trajectories & 8 to 16 \\ 
\hline
\end{tabular}
\label{GA_params}
\end{center}
\end{table}

The selection of the cost, \textit{i.e.}, fitness, function is often specific to the type of parameterization being evolved. The cost function selected in the present implementation is derived from the root-mean square (RMS) difference between the energy evaluations using the GA-derived parameter set and energy evaluations using DFT. Thus, we have 
\begin{equation}
\widehat{CF}(\mu,\text{basis set}) \, = \, \sqrt{{\frac{\displaystyle\sum_{i=1}^{N_\text{configs}} (E_{DFT}(\text{basis set}_i) \, - \, E_{GA}(\mu,\text{basis set}_i))^2}{N}}}
\label{RMS}
\end{equation}
where the subscript $GA$ indicates an energy evaluation using the shell model potential with a GA-derived parameterization and $N_\text{configs}$ is the total number of displacement configurations involved in the fitting. As a result, the GA strives to minimize the cost function of each parameter set and thus those population members with lower cost function values are of greater fitness.

\section{Application to a Shell-Model Potential for Metal-Oxides: BTO}

As an example, we apply the GA method to the parameterization of an interatomic potential for metal-oxides. The metal oxides are found in a diverse range of applications that include piezoelectrics transducers and actuators \cite{Sepliarsky2005, Iles2007}, ion carriers for batteries \cite{Kang2009}, catalysts for fuel-cells \cite{Yamamoto1986}, and elements for gas sensing and actuations \cite{Comini2006}, and yet currently there are not as established and tested interatomic potentials for metal-oxides as exist for the semiconductor or metal class of materials. This is possibly because the simulations of metal-oxides need to account for explicit charges (positive on the metal side and negative on the oxygen side) in the structures and during the dynamics. Therefore both non-bonding long range Coulomb and bonding short range interactions need to be properly incorporated in the potential.

To demonstrate the method and the advantages of using GA for optimizing a force field function over the traditional local gradient based methods, we have chosen to optimize the parameters of the shell model potential, which is suitable for non-reactive modeling of metal oxides in general and BTO in particular. Currently, numerous parameter sets exist in the literature for BTO type metal-oxides developed over the past years. Tinte et al \cite{Tinte1999} present a parameterization of the shell model potential using the phase transition properties in terms of lattice lengths and average electronic polarization to produce different crystal phases with increasing temperature. Sepliarsky et al \cite{Sepliarsky2005} have extended the potential to simultaneously model BTO as well as Barium Strontium Titanate (BST). Both of these studies review distinct facets of material behavior, but noteworthy is that  they offer different short-range parameters for the Buckingham potential and consequently strive to model different BTO properties.

Within the context of using the GA method for optimizing a BTO type metal-oxide, in the following we briefly describe (i) the ansatz or the functional form for the shell model potential for modeling metal-oxides, (ii) the structures and multiple phases of BTO which are needed as the underlying crystal structures for the input reference database, and (iii) the ab initio quantum mechanical DFT simulations for the structural (energy vs structure for different phases) and mechanical (changes in the energy vs structure) properties of BTO, which populate the reference dataset for the results shown in Section 4 of this work.

\subsection{Shell Model Potential for Metal-Oxides}

The shell model potential was initially proposed by Dick and Overhauser \cite{Dick1958} in order to capture the polarizability of dielectric materials within the context of a simplified potential field. Since its introduction and the advent of computational methods and systems, the shell model potential has been applied to an ever increasing range of dielectric materials \cite{Sepliarsky2005,Mitchell1993,Phillpot2007}. 

\begin{figure}[htbp]
\begin{center}
\includegraphics[scale=0.8]{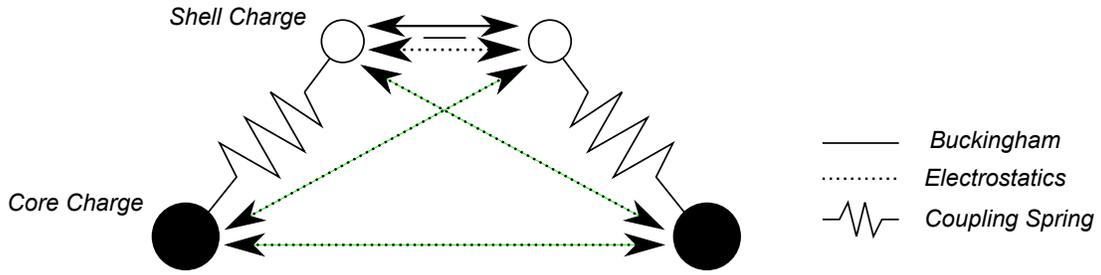}
\caption{Distinct interactions within the shell model potential}
\label{core-shell}
\end{center}
\end{figure}

The general premise of the shell model is to represent each ion in the system as a two-component entity: a positively-charged core and a negatively-charged shell which are coupled together through a spring-like potential. Long-range Coulomb, or electrostatics, effects are present between all the cores and shells of the system, except the core and shell of a given ion. Short-range Buckingham repulsion forces exist between the shells of neighboring ions up to a prescribed cut-off. In our model, Buckingham forces are present only between the shell of an oxygen and the shell of another ion. In addition, the core and shell of each ion are coupled by a spring-like potential consisting of harmonic and anharmonic components. The primary role of the coupling term is to affix each shell to its corresponding core, thus allowing the shell model potential to capture dipole-induced effects so as to predict the general polarizability of the dielectric. These interactions are summarized in \figref{core-shell}.

The general functional form of the potential is
\begin{equation}
\begin{split}	
U(\vect{r}, w) &= U_{\text{PME}}(q,\vect{r}) \, + \, U_{\text{BCKH}}(\vect{r}) \, + \, U_{\text{CPL}}(\vect{r}) \\ 
&= \frac{1}{2} \sum_n^{'}\sum_{i,j, i \neq j} \frac{q_i \, q_j}{|\vect{r}_{ij}+n|} + \sum_{\substack{ {i,j},{i \neq j} \\ |\vect{r}_{ij}| \le r_c }} \left( A \: \exp \left(-\frac{|\vect{r}_{ij}|}{\rho}\right)  - \frac{C}{|\vect{r}_{ij}|^6} \right) \\
& \ \ \ \, + \sum_i \, \left(\frac{c_2 \, w_i^2}{2} + \frac{c_4 \, w_i^4}{24}\right)
\end{split}	
\label{general_form} 
\end{equation}
The first summation term is the electrostatic contribution, the second is the short-range Buckingham potential, and the final term is the coupling term between the shell and core of a given ion. To avoid cumbersome notations, we do not precisely specify the particle pairs involved in each interaction. Instead all interactions that need to be considered are described in \figref{core-shell}. We denote
\begin{equation*}
	\vect{r}_{ij} = \vect{r}_i \, - \, \vect{r}_j	
\end{equation*}
the displacement vector between two particles, and
\begin{equation*}
	w_i = |\vect{r}_{\text{core of ion i}} - \vect{r}_{\text{shell of ion i}}| 
\end{equation*}
where $\vect{r}_{\text{core}}$ and $\vect{r}_{\text{shell}}$ are the coordinates of the core and shell of the same ion. Details of each contribution are described below.

For an ionic material, long range electrostatic forces are a vital contribution to the energy landscape of the dielectric system. A highly accurate and computationally efficient technique by which to treat electrostatic effects is the Particle Mesh Ewald (PME) method. The formulation of PME as applied here is based on Refs \cite{Essmann1995,Deserno1998,Frigo1997}. 

As noted previously, the coupling term $U_{\text{CPL}}(\vect{r}) $ in itself is a constraining force which allows the shell model to capture polarization within the dielectric's bulk crystalline structure by introducing a restoring force on the shell of a given ion. In effect, unlike the electrostatic or Buckingham short-range interactions, it is not directly correlated to the configurational energy of the basis geometry and thus the coupling constants are primarily adjusted to capture the correct polarizability of the dielectric. It is noted that in its initial form in \cite{Dick1958}, the coupling term consisted of only the harmonic constant $c_2$. However, in later iterations of the model, the anharmonic term $c_4$ was added to account for the effects of hybridization of the ionic structure \cite{Sepliarsky2005}.

For the current cost function as given in \eqsref{RMS}, our primary focus in this initial work is in accurately capturing the configurational energy landscape of the BTO basis in each of its four phases. Therefore, as a first approximation, we do not directly address the polarizability of the dielectric and the fitting of the harmonic and anharmonic terms of the coupling potential. The shell positions are then held fixed in relation to their respective cores, so that coupling potential remains consistent and constant through all four phases. In general, it is possible to modulate the Born effective charges \cite{Tinte1999} for the polarizability values under an applied field within the DFT framework \cite{Gonze2005}, and use those in fitting the coupling terms as well using the GA approach. 

The total parametric space of the shell model is now clearly identified with six distinct charges for the core and shell of each ion species, the same number of harmonic/anharmonic constants to define the coupling terms, and nine distinct coefficients for the Buckingham potential's oxygen-centric short range interactions. An overall summary of the interaction parameters is given in \tableref{Parameters}.

\begin{table}[htbp]
\begin{center}
\caption{Parameterization of the Shell Model Potential as Applied to BTO}
\begin{tabular}{|l| c c c c|}
 \hline
Potential & Parameters & Entity 1 & Entity 2 & Total $\#$ Parameters \\ \hline
Electrostatics & $q_{\text{core}}$, $q_{\text{shell}}$ & Core/Shell & Core/Shell & 6 \\ 
Buckingham & $A$, $\rho$, $C$ & Shell & Shell & 9 \\ 
Coupling & $c_2$, $c_4$ & Core & Shell & 6 \\ \hline \hline
\textbf{Total} &  & &  & 21 \\ \hline
\end{tabular}
\label{Parameters}
\end{center}
\end{table}

In effect, there are a total of 20 independent parameters, since the BTO basis has a neutral net charge, i.e., by prescribing five of the six system charges, the sixth charge directly results. 

\subsection{Crystal Structures and Structural Displacements of BTO}

\begin{figure}[htbp]
\begin{center}
 \subfigure[The basis for the cubic phase of BTO]{\label{cubic_structure}\includegraphics[scale=0.3]{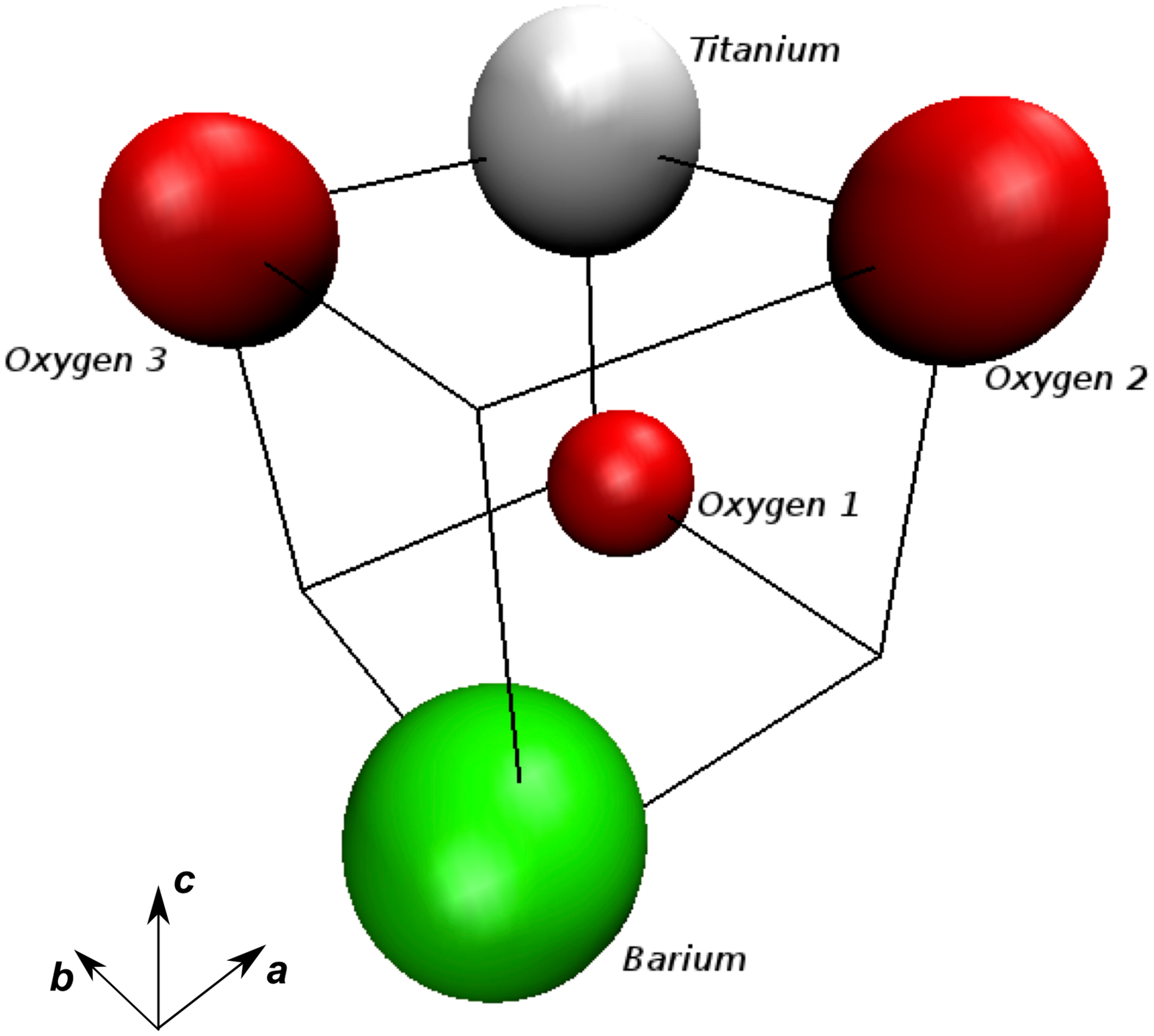}}
  \ \ \ \ 
\subfigure[Barium-Oxygen$_1$ type \textbf{3} displacement.]{\label{displacement}\includegraphics[scale=0.32]{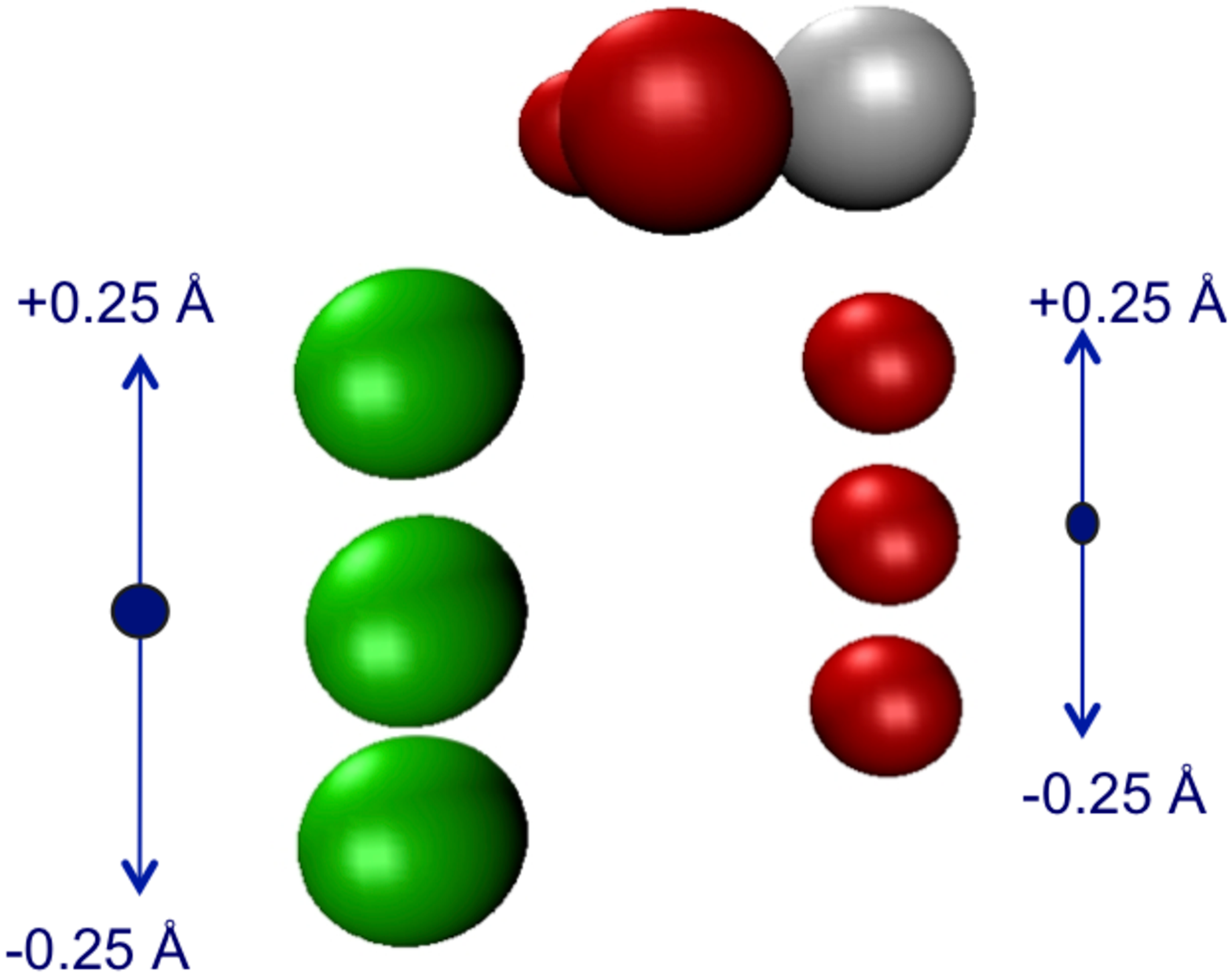}} 
\caption{Cubic basis and sample atomic displacement. Atom displacements begin at $-0.25$ \AA\ from their cohesive configuration and are incremented in 0.05 \AA\ steps.}
\label{cubic_phase}
\end{center}
\end{figure}

Our objective is to prepare a reference dataset including the lattice constants for original crystal structures and the displacements around their equilibrium configurations to represent the elastic properties of BTO in all four phases. We start by selecting the reference equilibrium configurations for each of the crystal phases and then proceed to permute the positions of each atom or pairs of atoms in turn where the DFT energy is determined for each new configuration.  The original configuration and the lattice constants for the cubic phase are shown in \figref{cubic_structure}.

All possible displacements of pairs of atoms along x, y, and z directions are considered. For any pair of atoms, this leads to 9 possible combinations of displacements. Each atom in a pair is moved in increments of 0.05 \AA\ up to a maximum of ±0.25 \AA\ from the equilibrium positions as determined in \cite{Kwei1993}. The different displacement orientations are listed in Table \tableref{disp}. An example of Type 3 displacement in \tableref{disp} is shown in \figref{displacement}.

\begin{table}[htbp]
\begin{center}
\caption{Types of Displacements of the BTO Basis}
\begin{tabular}{|c| c|}
 \hline
Displacement Key & Type of displacement \\ \hline
\textbf{1} & Ion 1 in the $\vect{a}$  direction / Ion 2 in the $\vect{a}$ direction \\
\textbf{2} & Ion 1 in the $\vect{b}$  direction / Ion 2 in the $\vect{b}$ direction \\
\textbf{3} & Ion 1 in the $\vect{c}$  direction / Ion 2 in the $\vect{c}$ direction \\
\textbf{4} & Ion 1 in the $\vect{a}$  direction / Ion 2 in the $\vect{b}$ direction \\
\textbf{5} & Ion 1 in the $\vect{a}$  direction / Ion 2 in the $\vect{c}$ direction \\
\textbf{6} & Ion 1 in the $\vect{b}$  direction / Ion 2 in the $\vect{c}$ direction \\
\textbf{7} & Ion 1 in the $\vect{b}$  direction / Ion 2 in the $\vect{a}$ direction \\
\textbf{8} & Ion 1 in the $\vect{c}$  direction / Ion 2 in the $\vect{a}$ direction \\
\textbf{9} & Ion 1 in the $\vect{c}$  direction / Ion 2 in the $\vect{b}$ direction \\ \hline
\end{tabular}
\label{disp}
\end{center}
\end{table}

In the phases with orthogonal primitive vectors, (i.e., orthorhombic, tetragonal and cubic phases), the orientation of the basis vectors $\vect{a}$, $\vect{b}$, and $\vect{c}$ correspond to the Cartesian axis and are dealt as above. The rhombohedral phase has non-orthogonal primitive vectors and requires the use of reduced coordinates to define its displacements consistently. Using a similar convention as defined in \cite{Laboratory}, the rhombohedral displacements are carried out along the appropriately defined primitive vectors, resulting in a maximum magnitude of displacement from the equilibrium position of $\pm$ 0.2 \AA\ per ion displaced.

It is noted that in this work, a comprehensive set of displaced basis configurations along the primary axes has been included in the reference database so as to ensure a complete resolution of the elastic properties of the material. In general, however, this may prove to be superfluous. A smaller dataset for near-equilibrium and far-from-equilibrium displacements with systematic, Boltzmann, and randomly distributed configurations could in fact prove to be sufficient for the reference dataset, and so many time consuming DFT structures and their respective energies may not be needed. This will be tested in further development of the methodology in the future.

To run the GA loop for the optimization, for a given set of the input parameters, the shell model energies described above are calculated for all the configurations listed above, whereas, the reference energies dataset comes from the ab initio quantum DFT simulations described later.  For the GA to efficiently calculate the shell model energies, a basis-replication scheme has been implemented within the optimization code. In this adaption, derived from conventional periodic boundary conditions \cite{AllenM.P.Tildesley1987}, images of the original basis are replicated in the surrounding space using the primitive translation vectors of the given phase. The interaction energies of the reference cell are summed up with energies of the images as well. A series of studies were used to determine the convergence in terms of the short range Buckingham cut-off, the PME cut-off, and the required number of replica images. Based on these studies, the short range cut-off was set at 10 \AA\ and a long range cut-off was set at 16 \AA. Further details of the image replication scheme are given in Solomon \cite{Solomon2012}.

Because of the symmetries in the crystal, many of the displacements actually produce configurations that are identical up to rotations and reflections. As a result, of the 90 total sets of displacements for the cubic phase, only 18 curves are needed to reproduce the required energy landscapes. The tetragonal phases has less intrinsic symmetries and therefore 48 distinct curves were required. Smaller subsets of net displacements were used for the two lower temperature phases, with the orthorhombic and rhombohedral phases referencing to 40 and 10 distinct displacement curves respectively. A summary of the reference displacement curves is given in \tableref{d_summary}.

\begin{table}[htbp]
\begin{center}
\caption{Reference Database of Displacement Curves \dagger}
\begin{tabular}{ l|c|c|c}
\hline
Phase & $\#$ Ions in Basis & $\#$ Displacement Curves & $\#$ Distinct Basis Configurations \\ \hline
Cubic        & 5  & 18 & 181 \\
Tetragonal   & 5  & 48 & 481 \\
Orthorhombic & 10 & 40 & 401 \\
Rhombohedral & 5  & 10 & 101  \\ \hline \hline
\multicolumn{3}{r|}{Total Configurations} & 1164 \\ \hline
\end{tabular}
\label{d_summary}
\end{center}
\begin{tablenotes}
\small
 \item \dagger These configurations comprise a full set of atomic pair displacements for the tetragonal and cubic phase up to rotations and reflections.
\end{tablenotes}
\end{table}

\subsection{DFT Energies of the Crystal Structures and Structural Displacements}

The DFT calculations were performed using ABINIT \cite{Gonze2005} in conjunction with ultra-soft pseudopotentials as prescribed by Vanderbilt \cite{Vanderbilt1990}. The differences in the symmetry of each phase require distinct kinetic energy cut-offs of the plane wave expansion as well as the distinct gridding of the reciprocal space. The key parameters of each series of DFT calculations were derived from a set of convergence studies. For cubic and tetragonal phases, energy cut-off's of 35 and 50 Ha respectively were used, with high density Monkorst-Pack gridding for the reciprocal space \cite{Monkhorst1976}. The orthorhombic phase has twice the number of ions in its basis at ten ions and as a result far less planes of symmetry in its geometry, thus requiring a higher energy cut-off of 70 Ha and a $6\times6\times6$ uniform reciprocal space grid. The rhombohedral phase with its non-orthogonal basis vectors required a high energy cut-off as well at 80 Ha, with a $8\times8\times8$ uniform reciprocal space grid.

\begin{figure}[htbp]
\begin{center}
\includegraphics[scale=0.45]{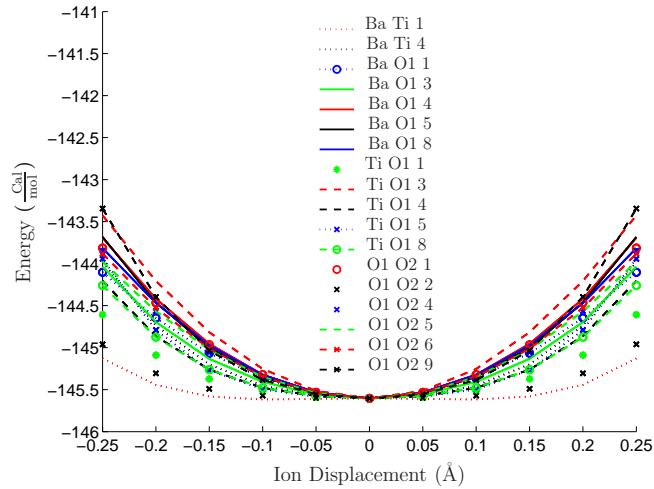}
\caption{Subset of DFT displacement curves for the cubic phase}
\label{dft_curves}
\end{center}
\end{figure}

The equilibrium configurations from which coupled-ion displacements are referenced were obtained from temperature dependent Neutron diffraction data presented by Kwei et al \cite{Kwei1993} used in conjunction with the reduced coordinate data given by Uludogan and Cagin \cite{Uludogan2006}. For the cubic phase, a relaxation of the ion structure was performed using a Broyden-Fletcher-Goldfarb-Shanno (BFGS) minimization cycle, rendering negligible change in the basis geometry. For the remaining phases, no minimization was attempted and the empirical lattice geometry provided by Kwei was used directly since the focal interest of the GA is the relative energy of displacements around the established equilibrium configurations. \tableref{basis_geo} presents the lattice geometry in the equilibrium configuration used of each of the four phases, and a sample of DFT energies corresponding to the cubic crystal phase and the displacements around the equilibrium configuration are shown in \figref{dft_curves}.

\begin{table}[htbp]
\begin{center}
\caption{Lattice Geometry of the Equilibrium Configurations}
\begin{tabular}{ |l|c|c|}
\hline
Phase & Primitive Vectors & Basis Angle \\ \hline
\ & \ & \ \\
Cubic        &  $\vect{a} \, = \, \left [ \begin{array}{c c c} 4.0 & 0 & 0 \end{array} \right ] $, $ \vect{b} \, = \, \left [ \begin{array}{c c c} 0 & 4.0 & 0 \end{array} \right ]$, & $\bot$ \\ \ & $ \vect{c} \, = \, \left [ \begin{array}{c c c} 0 & 0 & 4.0 \end{array} \right ]$   & \  \\
\ & \ & \ \\ \hline
\ & \ & \ \\
Tetragonal    &  $ \vect{a} \, = \, \left [ \begin{array}{c c c} 3.99095 & 0 & 0 \end{array} \right ] $, $ \vect{b} \, = \, \left [ \begin{array}{c c c} 0 & 3.99095 & 0 \end{array}\right ]$, & $\bot$ \\ \ &  $ \vect{c} \, = \, \left [ \begin{array}{c c c} 0 & 0 & 4.0352 \end{array} \right ]$   & \ \\
\ & \ & \ \\ \hline
\ & \ & \ \\
Orthorhombic   &   $ \vect{a} \, = \, \left [ \begin{array}{c c c} 3.9841 & 0 & 0 \end{array} \right ] $, $ \vect{b} \, = \, \left [ \begin{array}{c c c} 0 & 5.6741 & 0 \end{array} \right ]$, & $\bot$ \\ \ & $ \vect{c} \, = \, \left [ \begin{array}{c c c} 0 & 0 & 5.6916 \end{array} \right ]$   & \ \\
\ & \ & \ \\ \hline
\ & \ & \ \\
\ &   $ \vect{a} \, = \, \left [ \begin{array}{c c c} 3.26473 & 0 & 2.31843 \end{array} \right ] $, \ & \\ Rhombohedral   & $ \vect{b} \, = \, \left [ \begin{array}{c c c} -1.63237 & -2.82734 & 2.31843 \end{array} \right ]$, & $89.836^\text{ o}$ \\ \ & $ \vect{c} \, = \, \left [ \begin{array}{c c c} -1.63237 & 2.82734 & 2.31843 \end{array} \right ]$   & \  \\
\ & \ & \ \\ \hline
\hline 
\end{tabular}
\label{basis_geo}
\end{center}
\end{table}

\section{Results on the GA Optimization of BTO Interatomic Potential}

The GA methodology described above is applied to the tunable dielectric BTO, which is known to have four known stable phases or crystal structures in a rather small temperature range of a few hundred degree Kelvin. In the following, we describe the results of GA parameterization runs of the BTO potential using first an incomplete reference dataset comprising of DFT energies for the configurations of only two of the four crystal phases, followed by a second GA optimization in a single run with a complete reference dataset including DFT energies for all four of the crystal phases. The measure of accuracy of the first GA run with incomplete reference dataset with only two phases, cubic and tetragonal, is assessed by computing and comparing the structures and energies of the other two phases, rhombohedral and orthorhombic, which were not included in the incomplete reference dataset with their respective DFT calculated values. The accuracy of the second GA run with the complete dataset with all four crystal phases is assessed by the MD simulations of the basic elastic and thermal properties and comparing with the experimental and DFT results available in the literature. Additionally in this section we also show that, depending on the starting initial conditions, a conjugate gradient method for the same fit for a cubic phase has a tendency to get stuck in nearby local minima of the starting configuration, where as the GA method can accommodate a wide range of both near-equilibrium and far-from-equilibrium starting initial conditions or configurations in the fitting procedure.

\subsection{GA Parameterization Using Two BTO Phases: Incomplete Reference Dataset}

Using the data from a set of cubic and tetragonal DFT curves listed in \tableref{d_summary} (a total of 662 configurations), a series of four iterative GA evolution cycles were generated based on the higher temperature phases. In general, in the GA optimization procedure, the initial starting parameters in the first chromosomes are chosen randomly within a range defined by physics-based considerations for a given model (the shell model in our case). However, for our test case of BTO, a shell model potential with a parameter set (fitted through conventional techniques) has already been reported in the literature by Sepliarsky et al \cite{Sepliarsky2005}. This potential was used as a loose guide to choose the ranges within which the initial parameters in the starting chromosome were chosen randomly. Each individual parameter's initial regime was iteratively adjusted so that with all other parameters fixed, the parameter was able to explore a resulting energy range to within $\pm$ 100 kCal/mol. For a completely new case, similar criteria for a range of variations for the initial energies could be used to set-up the initial ranges for choosing the parameters randomly within the range.

\begin{figure}[htbp]
\begin{center}
\subfigure[Most fit parameterization: $1^{st}$ GA cycle --- per ion RMS: 0.4132~KCal/mol ]{\label{cubic_vs_dft}\includegraphics[scale=0.425]{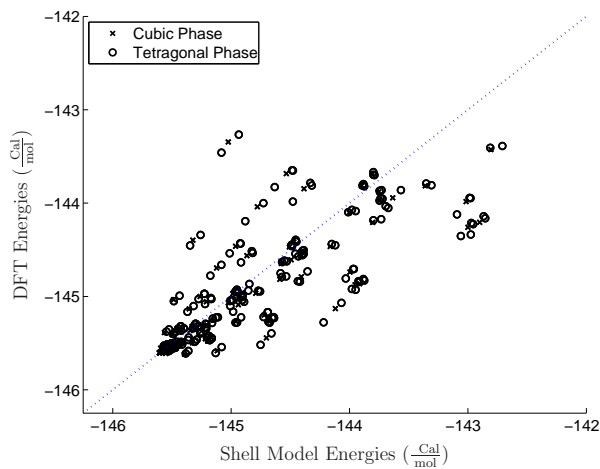}}
  \ \ \ \ 
\subfigure[Most fit parameterization: $2^{nd}$ GA cycle --- per ion RMS: 0.2366~KCal/mol]{\label{tetra_vs_dft}\includegraphics[scale=0.425]{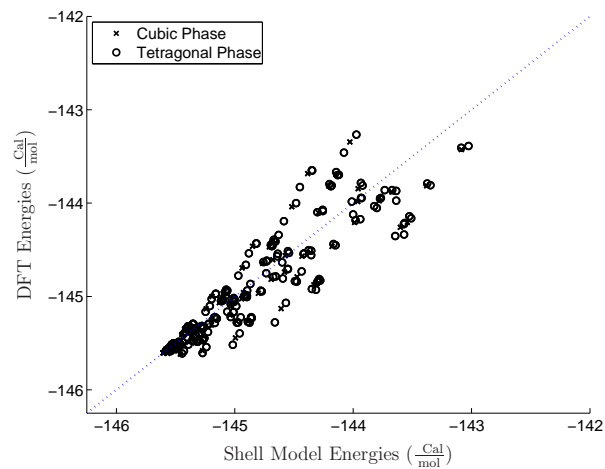}} 
\\
\subfigure[Most fit parameterization: $3^{rd}$ GA cycle --- per ion RMS: 0.1724~KCal/mol]{\label{ortho_vs_dft}\includegraphics[scale=0.425]{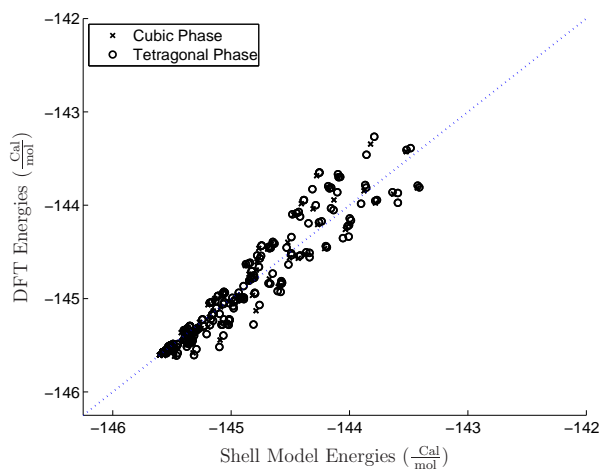}}
  \ \ \ \ 
\subfigure[Most fit parameterization: $4^{th}$ GA cycle --- per ion RMS: 0.1273~KCal/mol]{\label{rhom_vs_dft}\includegraphics[scale=0.425]{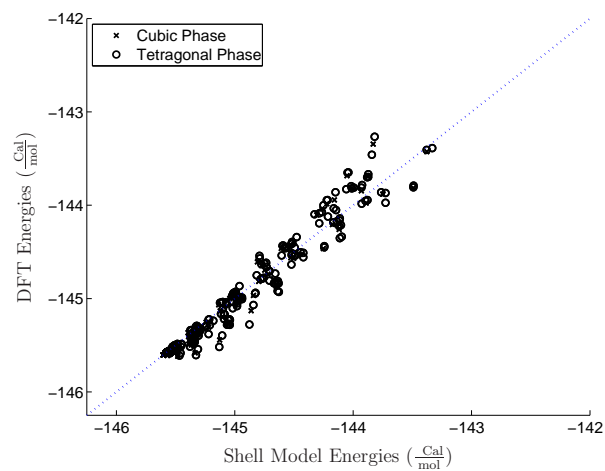}} 
\caption{DFT to GA shell model energy comparisons based on parameterizations derived from the GA evolutionary cycle. The dashed line here represents $y=x$; i.e., exact agreement. Plots illustrate the GA optimization cycle starting from the initial literature-derived parameterization space of the 1$^{st}$ GA cycle to an entirely GA redefined parameterization space which produces the most-fit parameter set of 4$^{th}$ GA cycle.}
\label{dft_vs_md}
\end{center}
\end{figure}

Once an initial evolution of 400 generations was created for 16 independent GA runs, the trajectories of each of the distinct parameters of the fittest member of the 16 different populations were studied. The GA allows for the definition of limits for each evolvable parameter. Those parameters which drifted to the edges or limits of their pre-defined range, (the $\eta_{min}$ and $\eta_{max}$ discussed in Section \ref{GA_section}), were allowed to explore the expanded range by shifting the limits during the subsequent evolution cycles. It is noted that no parameter's search range was ever contracted.  A comparison of the thus GA-fitted or optimized shell model energies with the input DFT energies in the reference dataset for the cubic and tetragonal phases, fitted simultaneously in a single GA optimization run over four cycles is shown in \figref{dft_vs_md}. In the current context, an evolution cycle consists of the 400 generations between expansion of the parameter exploration limits.

Overall improvement in the resulting parameterization is evident as the distribution of shell model and DFT energy data points start to cluster tightly around the $slope = 1$ line representing a perfect fit for all data points in the considered energy range. Unless otherwise noted, the RMS of each GA cycle is given on a per-ion basis so as to directly compare the quality of the fidelity across the different phases, specifically given that the orthorhombic basis consists of twice the number of atoms or ions. It is noted that as the GA optimization proceeds, the initial good fit is achieved at the low or lowest energy configurations even in the 1$^st$ GA cycle, which could be similar to local gradient based fitting methods focused entirely around the near-equilibrium configurations. We note that the RMS spread of the fitted and the input data-points for explicitly far-from-equilibrium (higher energy) configurations also start to decrease sharply in the subsequent GA cycles 2, 3, and 4 in the same single GA run. The procedure is stopped when the RMS of the spread of all the data-points for all the near-equilibrium and far-from-equilibrium configurations or structures is within some prescribed or accepted RMS value.

Another key differentiation with the traditional local gradient based methods is that in the GA fitting procedure the entire DFT based reference dataset is used in the fitting procedure and not any derived physical properties such as the elastic constants, which by definition are also limited to including only small deviated or displaced configurations from the equilibrium values. This brings out two key advantages of the GA fitting procedure over the typical local gradient based methods: (i) since the GA includes the entire raw data of the DFT structures and their respective energies in a single run it will be straightforward to incrementally include the additional data, such as DFT structures and energies of the additional phases, surfaces, and clusters as it becomes available, and repeat the GA run over a much shorter period of time than that required for ad-hoc fitting procedures to improve or extend the fitted potential, (ii) since both the near-equilibrium (low energy) and far-from-equilibrium (high energy) configurations or structures are explicitly included, the GA will be better suited to fit potentials with explicitly very high temperature and high reactivity data and properties in the dataset, which are generally not explicitly included in the fitting procedures for the interatomic potentials thus far.

\begin{figure}[htbp]
\begin{center}
\includegraphics[scale=0.7]{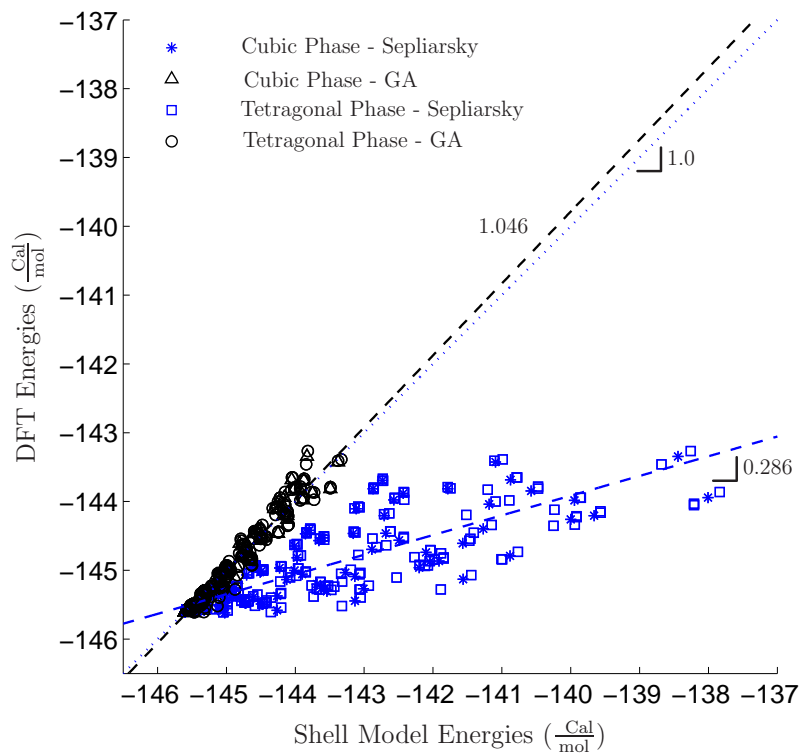}
\caption{Comparison of $4^{th}$ GA cycle parameterization to that of Sepliarsky et al \cite{Sepliarsky2005}. Plots that have linear fits, such as the one presented here, have the symbol $\righthalfcup$ to indicate a trend line. The plot emphasizes the role of GA optimization in producing a parameterization with greater fidelity to the ab initio calculations than that of current parameterizations presented in the literature.}
\label{ga_vs_sepl}
\end{center}
\end{figure}

As an example, in \figref{ga_vs_sepl} we compare the energies obtained from Sepliarsky et al \cite{Sepliarsky2005} parameterization for all the structures included in our dataset and our GA-fitted energies for the same structures from the parameterization after the 4$^{th}$ GA cycle, directly to the DFT energies for the same structures in the entire range.  For the near-equilibrium or lowest energy structures there is a good overlap between our GA-fitted values and the Sepliarsky's values with the corresponding DFT simulated structures and energies. However, as we include the far-from-equilibrium or higher energy data-points and compare the slopes of linear fits in the two cases, the slope of GA-fitted energies is 1.05 for nearly a one-to-one correspondence, whereas for the Sepliarsky's values not only does the slope of the linear fit to the data of 0.29 have a larger deviation from the ideal value of 1 but the RMS spread of energies around the linear fit is also much larger ($\sim$ 2.1 KCal/mol). This was expected as the traditional local gradient based methods generally focus on using only the near-equilibrium structures and energies in the fitting procedure. The evolved parameterization of the 4$^{th}$ GA cycle which produced the energy evaluations in \figref{ga_vs_sepl} is given in \tableref{GA_4th_Cycle}.

\begin{table}[htbp]
\caption{Parameterization of the Shell Model Potential after $4^\text{th}$ GA Cycle. For direct comparison to other parameterization given in the literature \cite{Shimada2008, Tinte1999, Sepliarsky2005}, units are given for length, charge and energy in terms of \AA, $e$ and eV.} 
\label{GA_4th_Cycle}
\begin{center}
\begin{tabular}{|l| c c | c c | c c c|}
\hline
Species & $q_{\text{core}}$ & $q_{\text{shell}}$ & $c_2$* & $c_4$* & $A$ & $\rho$ & $C$ \\ \hline
Ba	& 5.63546  & -3.66155 &	230.5152 & 107.6093  &	1721.4643 & 0.3347  &  43.8520 \\
Ti	& 3.62644  & -2.44219 &	304.8866 & 483.5996  &	2362.5255 & 0.2566  &  69.0100 \\
O	& 1.27176  & -2.32448 &	 29.4573 & 3952.6778 &	4179.7247 & 0.2768  & 107.8916 \\ \hline 
\end{tabular}
\small{\textit{*Un-evolved parameter.}}
\end{center}
\end{table}

\begin{figure}[htbp]
\begin{center}
\subfigure[GA prediction of orthorhombic phase --- per ion RMS: 0.1933~KCal/mol]{\label{ortho_vs_dft}\includegraphics[scale=0.39]{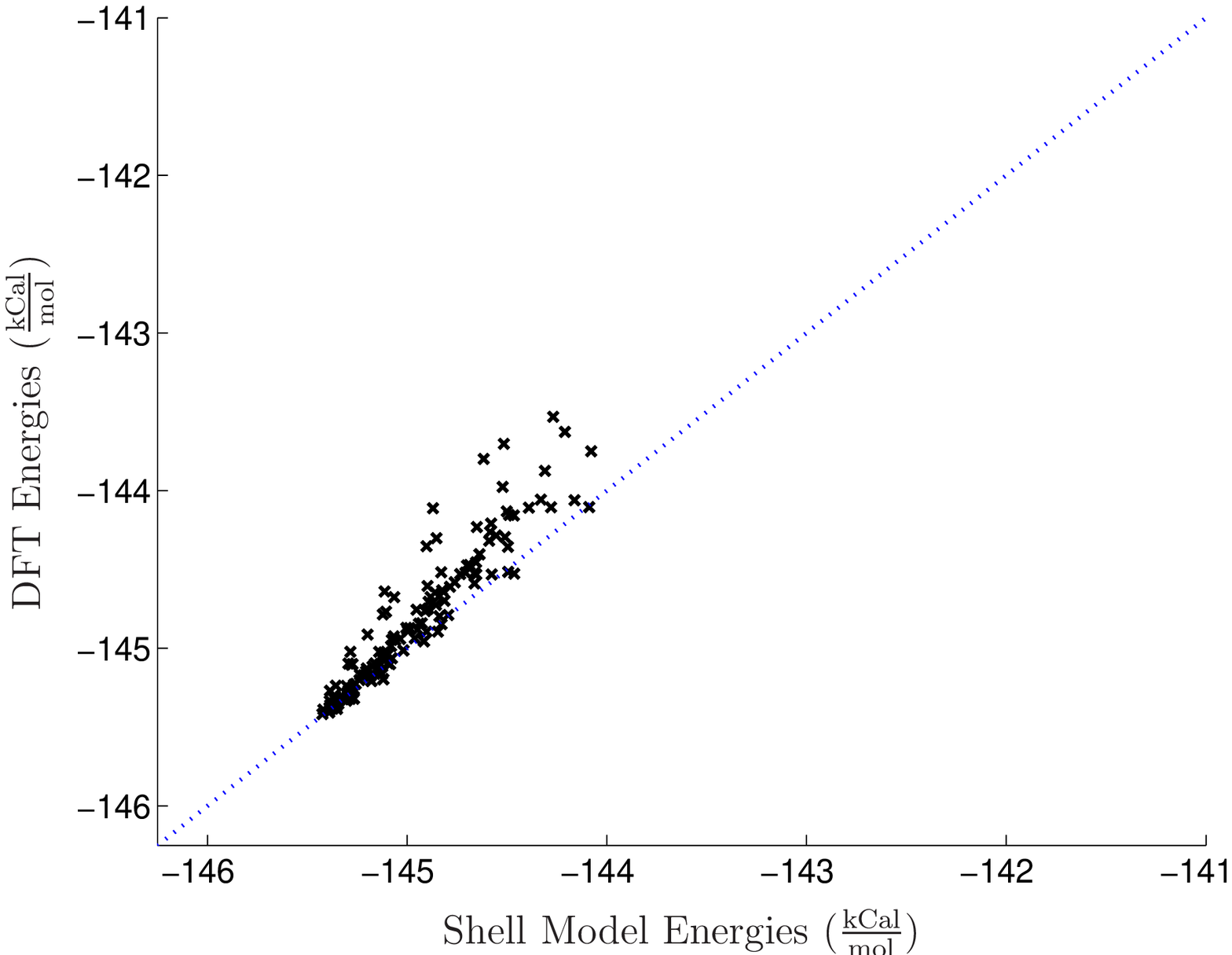}}
  \ \ \ \ 
\subfigure[GA prediction of rhombohedral phase --- per ion RMS: 0.1599~KCal/mol] {\label{rhombo_vs_dft}\includegraphics[scale=0.39]{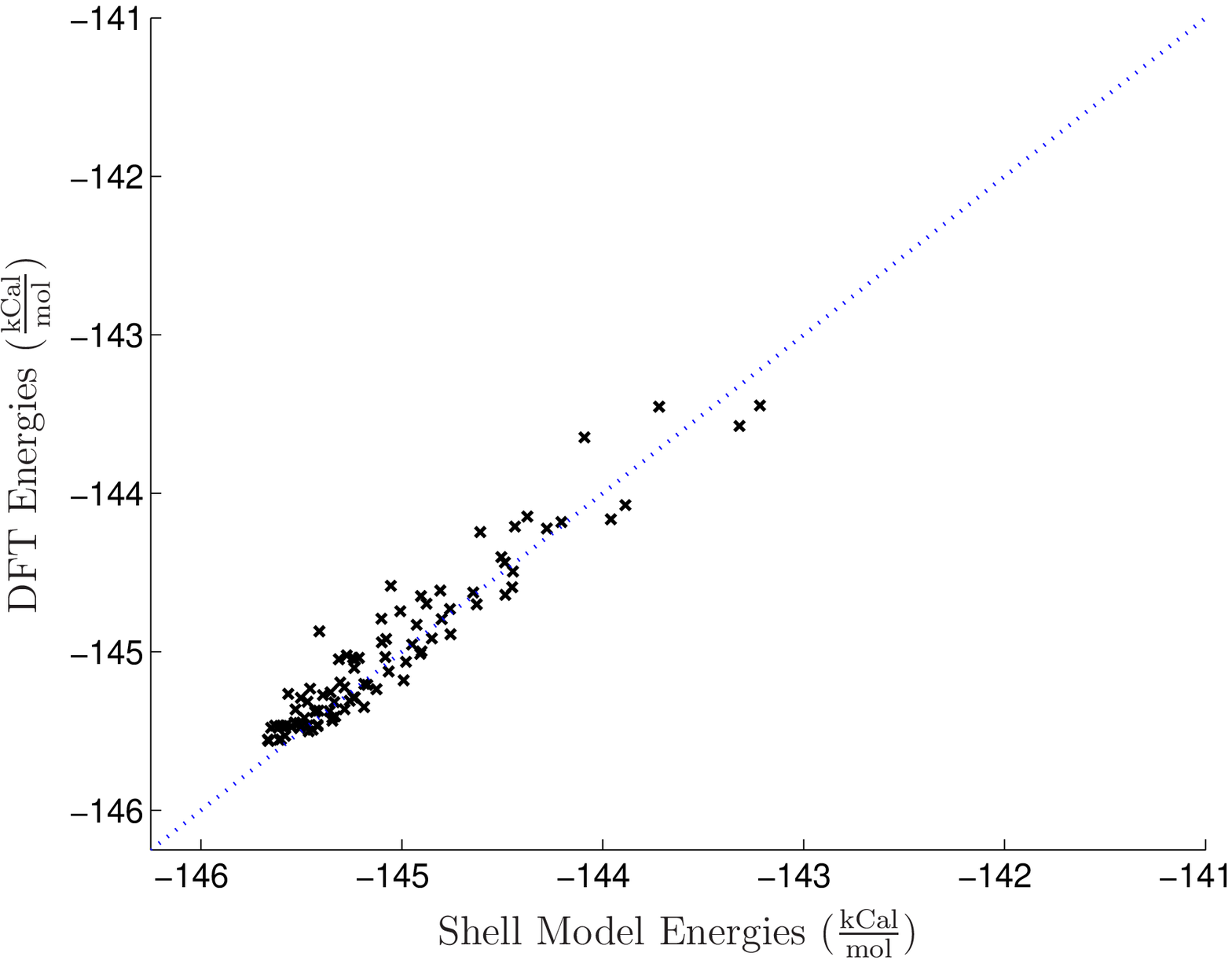}} 
\caption{GA predictive parameterization of the orthorhombic and rhombohedral phases as compared to the reference DFT data set}
\label{ga_predict}
\end{center}
\end{figure}

Lastly in this section we assess the robustness of the above GA optimized potential fitted with only the tetragonal and cubic phases of BTO in the incomplete reference dataset. This is done by predicting the near-equilibrium (low energy) and far-from-equilibrium (high energy) configurations of the other two orthorhombic and rhombohedral phases of BTO not included in the reference dataset. The GA-fitted energies for orthorhombic and rhombohedral phases using the incomplete reference dataset are directly compared to their DFT values simulated for the complete reference dataset to be used in the next section. The comparison is shown in \figref{ga_predict}. Interestingly, the comparison shows that the structures corresponding to rhombohedral phase are captured accurately in the entire energy range with an RMS between the predicted and DFT values to be 0.1599kCal/mol. The predicted values of orthorhombic phase show a larger deviation for higher energy structures but low energy structures are in good agreement.

These results show the utility and robustness of the method given that the potential for having to contend with a Pareto set scenario is greatly reduced. Pareto set analysis, \cite{Back1996, Eiben2007}, applies to evolution strategies with multiple cost functions in which no one solution dominates all others. The concept of Pareto front is central in problems where the cost functions are somehow in conflict with one another and therefore the optimal solution involves in general a trade-off amongst the competing objectives. The above results indicate that the simultaneous optimization for the two (cubic and tetragonal) phases is in relative congruence with an optimization for the remaining two (orthorhombic and rhombohedral) phases of the broader perovskites family of materials, which is an interesting result in itself and indicate that optimization for one phase does not occur at the detriment or cost of another. This may not be case in fitting other structures such as clusters, surfaces, and reactivity etc. In that case, with GA it is possible to also specify the relative importance of one desired set of properties over the others. In the later Section \ref{full_ga_section}, we will describe a simultaneous optimization with all four phases in the complete reference dataset.

\subsection{Comparison of GA Optimization with a Local Gradient Based Method}

One of the main limitations of the local gradient based methods, as used for the fitting of the complex interatomic potentials thus far, has been to fit one set of the properties such as bulk structure or crystal phases at a time. For an overall optimization for a system with multiple objectives, such as multiple crystal phases or properties, a serial procedure is often used because one phase or property is fitted at a time and one needs to manually go back and forth to iteratively fit the parameters in different sets to get overall averaged parameters for more than one crystal phase or property. That is why typically it has generally taken years to incrementally improve a previously optimized potential to incorporate additional data for new phases or properties.  In the case of BTO, we seek a single parameter set capable of accurately representing the lattice constants and mechanical properties across the four known phases. 

\begin{figure}[htbp]
\begin{center}
\includegraphics[scale=0.7]{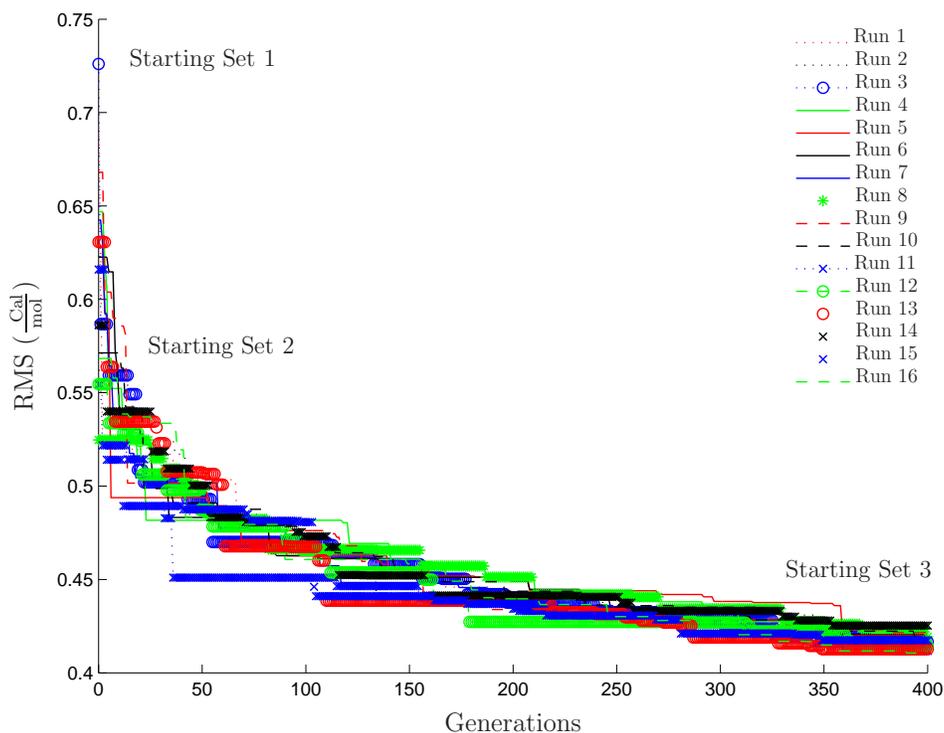}
\caption{Three distinct starting points for the CG minimization routine}
\label{cg_start}
\end{center}
\end{figure}

Because of the above limitation on the typical use of local gradient based methods to fit only one system or crystal phase at a time, in this section, we compare the GA approach with conjugate gradient (CG) technique applied only to the well known high temperature cubic phase data for BTO. The Brent method \cite{Press2002} was used in conjunction with the robust bracketing method to determine the step size in the line search direction. For validation, the method was initially applied to a number of single displacement curves, and for all curves tested, CG parameterizations with an RMS of $\sim$ 0.03 kCal/mol were obtained.  In applying the procedure to the full set of cubic phase curves, different parameter sets from a GA evolution of the cubic phase were selected as the starting points. \figref{cg_start} shows 16 distinct GA trajectories. The CG minimization was initiated from three distinct parameter sets along these trajectories. The results of three subsequent CG minimization trajectories from these three initial starting points are reported in \tableref{cg_minim}. In all cases, the CG initiates a minimization trajectory, but with varying degrees of success resulting in little to no improvement in the fidelity of the parameterization to the DFT energies.

\begin{table}[htbp]
\caption{GA Minimization Results}
\begin{center}
\begin{tabular}{|l| c  c|c|}
\hline
Starting Set & Initial Value & End Value & Reduction in RMS \\ \hline
Set 1	& 0.7260      &  0.7190    &	$\sim$ 0.007 \\
Set 2	& 0.5711      &	 0.5605    &    $\sim$ 0.0106 \\
Set 3	& 0.4106164   &  0.4106126 &	$\sim 3.8 \times 10^{-6}$ \\ \hline
\end{tabular}
\label{cg_minim}
\end{center}
\end{table}

\subsection{GA Parameterization Using All Four BTO Phases and Simulations of Materials Properties}
\label{full_ga_section}
This next set of results is derived from the GA evolution based on displacement data from all four phases of BTO. In this section, we first describe the GA optimization of the BTO potential with all four phases, followed by the simulation of the basic structural, mechanical, and thermal properties of BTO for comparison with experimental and simulation results reported in the literature.

\subsubsection{GA Parameterization with All Four Phases of BTO}

In this context, a subset of the total DFT database was selected for the fitting procedure so as to reduce the computational cost of the evolutionary cycle. Since different numbers of curves were used to represent each of the distinct phases, a weighted-sum fitness cost function was selected so that all phases would have an equivalent impact on the evolved parameterization. The general form of the weighted cost function is

\begin{align}
\widehat{CF}_{\text{Weighted}}(\mu,\text{basis set}) \, =& \, \sum_{j=1}^{N_{\text{Phases}}} \text{Weight}_{(j)} \, \times  \notag \\ &\sqrt{{\frac{\displaystyle\sum_{i=1}^{{N_{\text{Configs}_{(j)}}}} (E_{DFT}(\text{basis set}_i) \, - \, E_{GA}(\mu,\text{basis set}_i))^2}{N}}}
\label{weighted_sum}
\end{align}

\begin{figure}[htbp]
\begin{center}
\includegraphics[scale=0.7]{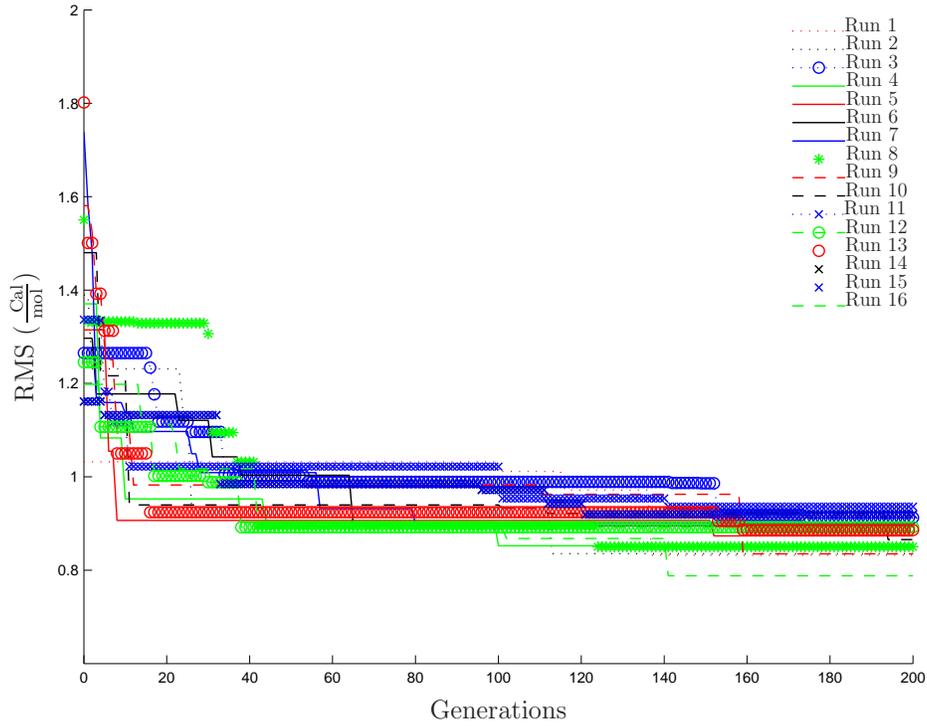}
\caption{GA evolution trajectories based all four BTO phases}
\label{GA_all_phases}
\end{center}
\end{figure}

\begin{figure}[htbp]
\begin{center}\includegraphics[scale=0.7]{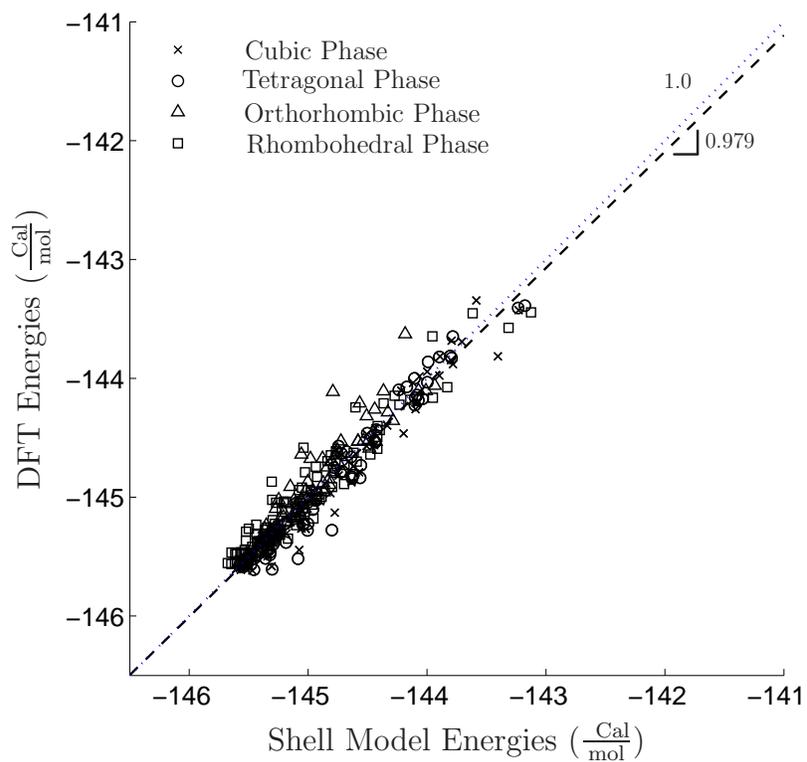}
\caption{DFT to GA evolved shell model energy comparison with linear fit to total phase spectrum; slope of linear fit $\sim 0.979$.}
\label{final_ga}
\end{center}
\end{figure}

The subset of DFT basis configurations used in the current evolution is given in \tableref{full_ga_table}. Reviewing the results of 16 distinct GA trajectories, it is found that a net weighted RMS $\sim$ 0.788 was achieved, as shown in \figref{GA_all_phases}. Plotting energies of all four GA-fitted phases in comparison with their respective DFT energies in \figref{final_ga}, we show a linear fit to the data (with slope $\sim$ 0.979) for both the near-equilibrium and far-from-equilibrium configurations or structures in the entire energy range. This shows a very good one-to-one agreement between the GA-fitted energies with their corresponding DFT energies for all the structures in the entire energy range. The GA evolved parameters corresponding to all four phases in the reference DFT dataset are given in \tableref{GA_Final_Params}.

\begin{table}[htbp]
\begin{center}
\caption{Subset of Displacement Curves for Full Spectrum Evolutions}
\begin{tabular}{ l|c|c|c}
\hline
Phase & $\#$ Ions in Basis & $\#$ Displacement Curves & $\#$ Distinct Basis Configurations \\ \hline
Cubic        & 5  & 18 & 181 \\
Tetragonal   & 5  & 18 & 181 \\
Orthorhombic & 10 &  9 & 91  \\
Rhombohedral & 5  &  9 & 91  \\ \hline \hline
\multicolumn{3}{r|}{Total Configurations} & 544
\end{tabular}
\label{full_ga_table}
\end{center}
\end{table}

\begin{table}[htbp]
\caption{Parameterization of the Shell Model Potential as Applied to BTO. Units are given for length, charge and energy in terms of \AA, $e$ and eV.}
\begin{center}
\begin{tabular}{|l| c c | c c | c c c|}
 \hline
Species & $q_{\text{core}}$ & $q_{\text{shell}}$ & $c_2$* & $c_4$* & $A$ & $\rho$ & $C$ \\ \hline
Ba	& 5.59664  & -3.14348 &	233.2602 & 254.3762  &	2489.3499 & 0.3281  &  41.4167 \\
Ti	& 3.41576  & -2.35313 &	291.2780 & 366.8786 &	2093.7480 & 0.2581  &  61.4168 \\
O	& 1.10895  & -2.28088 &	 19.1422 & 3839.1070 &	2469.9863 & 0.2658  &   5.6318 \\ \hline \hline
 & &  & \multicolumn{5}{c|}{Net Weighted RMS: 0.7879} \\ \hline
\end{tabular}
\small{\textit{*Un-evolved parameter.}} \label{GA_Final_Params}
\end{center}
\end{table}

\subsubsection{Simulations of Basic Structural, Mechanical, and Thermal Properties of BTO}

An assessment of the quality of the GA parameterization can be done by the simulations of the basic structural, mechanical, and thermal properties of BTO using the optimized parameters given in \tableref{GA_Final_Params}, and comparing the results with experimental or DFT simulated results reported elsewhere in the literature. We do that using both the static structural and dynamic MD simulation methods.

Using the equilibrium geometries given in \tableref{basis_geo}, a series of iterative adjustments to the unit cell were performed in each phase to arrive at the equilibrium lattice lengths and bulk modulus. At each unit cell adjustment, the energy of the specific configuration was evaluated using the shell model potential with the GA optimized parameters presented in \tableref{GA_Final_Params}. For the cubic and rhombohedral phases, the unit cell volume was adjusted by uniform expansion and contraction. However, in the tetragonal and orthorhombic phases, where lattice lengths are dissimilar, two sets of separate calculations were performed in each phase. First, to find the equilibrium lattice length each axis was adjusted independently. Second, for the bulk modulus calculation the axes were adjusted uniformly as in the cubic and rhombohedral phases. Fitting a fourth-order polynomial to the energy, and scaling by atomic volume, the bulk modulus was directly calculated by

\begin{equation}
\label{BM}
B = \text{V} \, \frac{\partial^2 \Phi}{\partial^2 \text{V}}
\end{equation}
where $\Phi$ is the equation of the polynomial. Performing this series of calculations, the cohesive volumes, equilibrium lattice and bulk modulus were calculated for each phase of the material, and compared with experimental or other simulated values reported in the literature. These values are presented in \tableref{statics}. As can be seen from the tabulated results, the GA parameterized potential accurately captures the physical properties in all four phases. The equilibrium lattice lengths and thus the stable unit cell volumes are in good agreement with the values reported in the literature. Similarly, the bulk modulus is also well represented throughout the phases and there is notable close agreement with cubic phase experimental values \cite{Pruzan2002}.

\begin{table}[htbp]
\begin{center}
\caption{Static Energetic Phase Characterizations}
\label{statics}
\begin{tabular}{|l| c c c|}
 \hline
\multicolumn{4}{|c|}{\bf{Cubic Phase}} \\ \hline
Source & Unit Cell Vol (\AA$^3$) & Equil. Lattice (\AA) & $B$ (GPa) \\ \hline
GA Evolved * & 63.426 & 3.988 & 124.17 \\ \hline
Experimental \cite{Pruzan2002} &  -     &  -   & 135 \\ \hline
Experimental \cite{Hellwege1969} & 64.00  & 4.00 & 162 \\ \hline
DFT \cite{Uludogan2006}          &  65.85 & 4.0382 & 160.84 \\ \hline
DFT \cite{Uludogan2002}          &  64.28 & 4.0058 & 167.64 \\ \hline 
MD  \cite{Tinte1999}		 &  64.77 & 4.016  & 226 \\ \hline
\multicolumn{4}{|c|}{\bf{Tetragonal Phase}} \\ \hline
Source & Unit Cell Vol. (\AA$^3$) & Equil. Lattice (\AA) & $B$ (GPa) \\ \hline
GA Evolved *                 & 63.121  & a = 3.9750, c = 3.9948 & 116.97 \\ \hline
Experimental \cite{Kwei1993} &  64.271& a = 3.99095, c = 4.0352  & - \\ \hline 
DFT \cite{Uludogan2006}      &  67.50 & a = 4.00480, c = 4.2087  & 82.94 \\ \hline
DFT \cite{Uludogan2002}      &  65.95 & a = 3.9759,  c = 4.1722  & 98.60 \\ \hline 
MD \cite{Tinte1999}          &  64.75 & a = 4.002,   c = 4.043   & - \\ \hline
\multicolumn{4}{|c|}{\bf{Orthorhombic Phase}} \\ \hline
Source & Unit Cell Vol. (\AA$^3$) & Equil. Lattice (\AA) & $B$ (GPa) \\ \hline
GA Evolved *              & 126.108 & a = 3.9682, b = 5.6287,  c = 5.6461   & 110.54 \\ \hline
Experimental \cite{Kwei1993} &  128.66 & a = 3.9984, b = 5.6741,  c = 5.6916  & - \\ \hline 
DFT \cite{Uludogan2006}      &  67.81 \dagger & a = 3.9914, b = 5.7830,  c = 5.8223    & 87.39 \\ \hline
DFT \cite{Uludogan2002}      &  66.02 \dagger & -    & 97.54 \\ \hline 
MD \cite{Tinte1999}          &  64.63 \dagger & a = 3.995, b = c = 4.022 \dagger & - \\ \hline
\multicolumn{4}{|c|}{\bf{Rhombohedral Phase}} \\ \hline
Source & Unit Cell Vol. (\AA$^3$) & Equil. Lattice (\AA) & $B$ (GPa) \\ \hline
GA Evolved *                 &  63.248 & a = 3.9842, $\alpha$ = 89.836 $^{\dagger}$ & 116.57 \\ \hline
Experimental \cite{Kwei1993} &  64.201 & a = 4.0042, $\alpha$ = 89.836  & - \\ \hline 
DFT \cite{Uludogan2006}      &  67.76  & a = 4.073, $\alpha$ = 89.74    & 94.62 \\ \hline
DFT \cite{Uludogan2002}      &  65.99  & a = 4.042, $\alpha$ = 89.77    & 103.5 \\ \hline 
MD  \cite{Tinte1999}         &  64.577 & a = 4.012, $\alpha$ = 89.81    &   -   \\ \hline
\end{tabular}
\small{\ \\ \textit{* Current work.}}
\small{\ \\ \textit{ \dagger This value is based on the pseudomonoclinic cell discussed by \cite{Kwei1993}.}}
\end{center}
\end{table}

As a next step of the validation of the GA evolved parameterization in \tableref{GA_Final_Params}, we use the optimized shell model parameters in MD simulations at 400K at which the BTO has been shown to exhibits the cubic phase in the experiments \cite{Kwei1993}. However, so far, with a focus on developing and showing the advantages on the GA method, we have applied the GA procedure only to the short range Buckingham and long range Coulomb potentials with spring constants effectively set to zero since shell positions are fixed to the respective core positions. In temperature dependent dynamic simulations, the spring constants for the core-shell coupling, however, cannot be zero and need to be selected.  Very small or soft spring constants may not apply enough restoring forces, whereas very large or stiff spring constants may lead to a neglect of the polarization effects observed in lower temperature phases for BTO.

The initial spring constants were thus selected in the medium range as reported in the literature and adjusted iteratively in a series of NVT and NPT simulations for the cubic phase at 400K such that the crystal structure during the dynamics remained stable and general polarization was close to zero.  The selected spring constants are listed in \tableref{k_set}. The following reported MD simulations were performed using the package DL POLY \cite{Smith2011} on a $10\times10\times10$ unit cell simulation domain with spring constants listed in \tableref{k_set}.

\begin{table}[htbp]
\begin{center}
\caption{Spring Coupling Term Set}
\label{k_set}
\begin{tabular}{| c | c  c |}
 \hline
Element & $c_2$ & $c_4$ \\ \hline
Ba & \ \ 900 \ \  & \ \ 0 \ \ \\ \cline{2-3}
Ti & \ \ 1170 \ \ & \ \  3600 \ \ \\ \cline{2-3}
O  & \ \ 500 \ \ & \ \ 4200 \ \ \\ \hline
\end{tabular}
\end{center}
\end{table}

\begin{figure}[htbp]
\begin{center}\includegraphics[scale=0.6]{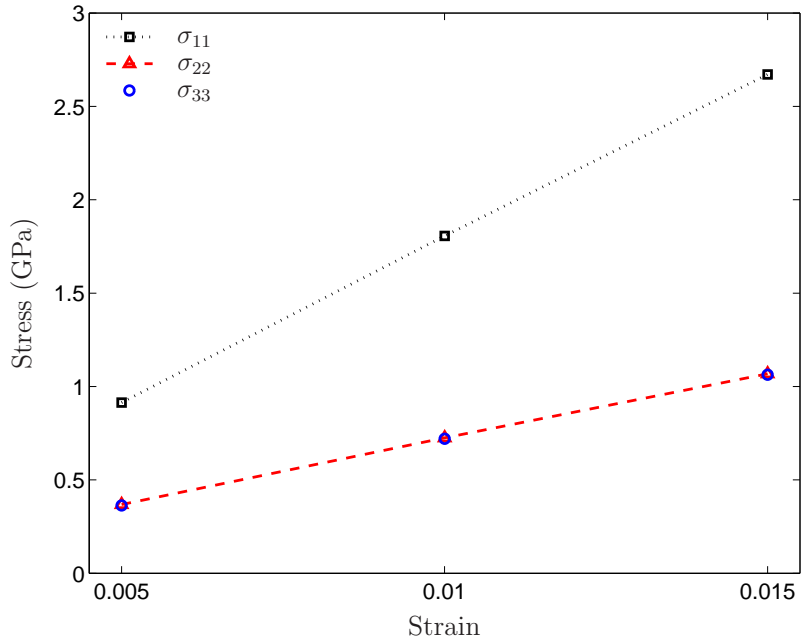}
\caption{Stress-strain relationships under the $\varepsilon_{11}$ strain condition at 400K}
\label{StressStrain}
\end{center}
\end{figure}

The mechanical and basic thermal properties were thus simulated in a series of MD simulations at 400K in the cubic phase using a $10\times10\times10$ sample of BTO. In addition to thermalization and lattice equilibration studies, the BTO block was loaded under 0.5\% to 1.5\% uni-axial strain along each primary axis to determine the elements of the stiffness tensor. For example, the diagonal components of the stiffness tensor as a function of strain at 400K are shown in \figref{StressStrain}. As a result of this series of simulations, the equilibrium lattice lengths, equilibrium volume, general polarization, uni-axial components of the stiffness tensor, the bulk modulus and coefficient of thermal expansion (CTE) were calculated at 400K, and summarized in \tableref{cubic_md}. Experimentally determined values, where available, are also given for comparison.

\begin{table}[htbp]
\begin{center}
\caption{Material Characterization at 400K}
\label{cubic_md}
\begin{tabular}{| c | r l | r l l |}
 \hline
Quantity      & \multicolumn{2}{|c|}{Current Work}  & \multicolumn{3}{|c|}{Reference Value} \\ \hline
$a$	      &    3.991 & \AA\	     & 4.00    & \AA\ & \cite{Hellwege1969} - Experiment \\ \hline
Equil. Volume &   63.560 & \AA$^3$           & 64.00   & \AA$^3$   & \cite{Hellwege1969} - Experiment \\ \hline
P$_x$         &   -0.410 &   & $\sim$ -2 &   & \\
P$_y$	      &   -0.247 & $\frac{\text{micro C}}{\text{cm}^2}$ & $\sim$ -1.75 & $\frac{\text{micro C}}{\text{cm}^2}$ $^*$ & \cite{Tinte1999} - MD \\
P$_z$         &   -0.761 &   & $\sim$ -1.75 & & \\ \hline 
C11	      & 182.769  & GPa  & - & & - \\
C22           & 183.170  &  & - & &  \\
C33	      & 182.172  &  & - & &  \\
C12 (C21)     &  73.325  &  & - & &  \\
C13 (C31)     &  72.813  &  & - & &  \\
C23 (C32)     &  73.068  &  & - & &  \\ \hline
$B$           & 109.614 & GPa & 135, 162 & GPa & \cite{Pruzan2002}, \cite{Hellwege1969} - Experiment \\ \hline
CTE	      & 9.879$\times$10$^{-06}$ & per $^o$C & 11.3$\times$10$^{-06}$ & per $^o$C & \cite{Bland1959} - Experiment \\ \hline
\end{tabular}
\flushleft{\small{\textit{*Values are approximated from plot on page 9687 of \cite{Tinte1999} at 220K, for which the reported parameterization exhibits the cubic phase.}}}
\end{center}
\end{table}

As can be seen from the values in \tableref{cubic_md}, the GA evolved parameterization captures the cubic phase with a high degree of fidelity while the various material properties are simulated. It is interesting to note that both the equilibrium volume and bulk modulus compare very well with experimental results, especially noting that the Pruzan et al \cite{Pruzan2002} value is of a more recent experimental work than that of Hellwege \cite{Hellwege1969}. In addition, the CTE is in good agreement with Bland results \cite{Bland1959}, indicating good correlation with experimental changes in the lattice constants as the temperature is varied around 400K. Finally, we also note that a set of input files has been prepared for the direct application of the parameterization on BTO in the \nst ensemble for the MD package DL\_POLY \cite{Smith2011}, and can be acquired by contacting J.E. Solomon.

\section{Summary on the Advantages of the GA Method}

In summary, we have set-up the background and framework for a GA method that can be used to systematically obtain and optimize parameters for interatomic force field functions for MD simulations by fitting to an energy landscape or reference database. As an example, we have developed and applied the GA method to the fitting of a potential for BTO, a member of an important class of materials metal-oxides, which have been known to have applications in a wide variety of areas such as tunable and low-k dielectrics, energy storage, fuel-cells, and automotive or environmental catalysts. The resulting GA optimized parameterization of the BTO potential has been able to (i) reproduce the relaxed structures and energies in good agreement with their corresponding DFT values for both the near-equilibrium and far-from-equilibrium configurations, (ii) capture the basic structural, mechanical, and thermal properties of the BTO up to 400K, and (iii) predict the two crystalline phases of the BTO, (orthorhombic and rhombohedral) which were not included in an incomplete reference DFT dataset simulation as a test case.

Using BTO as a testbed, the GA technique is shown to be able to contend with both the stiffness of the potential framework as well as the rugged topology of the system's energy landscape. In future work, where other materials of a more exotic nature are investigated, the methods key strengths of (i) global optimizations where multiple local minima may be present and (ii) the ability to add incremental DFT and experimental data for a diverse set of properties to the reference set with little to no modifications in the optimization procedure will prove valuable assets. For a given functional form, in the GA optimization procedure, the quality of the finally fitted force field function is limited only by the quality of the data in the reference dataset, which can be improved incrementally by both advanced experimental and simulation approaches.

In addition, the success of the GA method also stems from the single-objective framework in which the sole cost function, while being of relatively low computational cost, is able to reflect the quality of the parameterization in a wide range of physical configurations. From the perspective of the GA, the reference dataset is nothing more than a list of configurations and their respective energies. In actuality, these energies and configurations may represent a diverse landscape with respect to temperature, stress-strain, surfaces, clusters, and reactivity conditions, but the GA optimization procedure remains the same. By coupling the GA approach to wider and more robust databases in future, it may be possible to evaluate not only the existing functional forms for the accurate force-field parameterization for MD simulations, but also to search for new functional forms which are needed but are not available today.

\section{Acknowledgements}

This research was performed at Stanford University and supported in part by the U.S. Army Research Laboratory (ARL), through the Army High Performance Computing Research Center, Cooperative Agreement W911NF$-$07$-$0027. Additional funding for J.E. Solomon came from the Alfred P. Sloan Foundation and the National Science Foundation Graduate Research Fellowship. Special acknowledgement goes to L. Munday of ARL for his insightful talks with J.E. Solomon with regards to the use of \verb|DL_POLY|.

\newpage
\nocite{*}
\bibliographystyle{elsarticle-num} 
\bibliography{Method}

\end{document}